\documentclass{elsart}
\usepackage{t1enc,epsf}
\newcommand{\reffz}[1]{\textup{(\ref{#1})}}

\begin{document}

\begin{frontmatter}
\title {Detection of weakly interacting light vector bosons by coherent
scattering}
\author{A. Ljubi\v{c}i\'{c}\thanksref{email}}
and
\author{D. Kekez}
\address{Ru{\dj}er Bo\v{s}kovi\'{c} Institute, Bijeni\v{c}ka 54, 10000
Zagreb, Croatia}
\date{\today}
\thanks[email]{E--mail address: Ante.Ljubicic@irb.hr}

\begin{abstract}

We propose a novel experimental method to produce and detect
weakly interacting light vector bosons using coherent processes in
refractive media. A light vector boson would be produced by a
laser photon scattered at a plane interface between two media of
different dielectric properties and will be converted back to
photon in the similar scattering process. The effect depends
strongly on the indices of refraction of the two media. If
incident and recovered photons were traveling through extremely
high vacuums, limits on the existence of several light vector
bosons could be substantially improved.



\end{abstract}

\end{frontmatter}

\newpage

The existence of weakly interacting light bosons could be of
considerable significance in improving our understanding of many
phenomena. For example, such pseudoscalar particles as axions [1]
and majorons [2], could make contributions to the missing mass of
the universe and could also be involved in explanations of
possible neutrino masses. Hypothetical spin 1 baryonic [3] and
leptonic [4] photons are candidates to be the quanta of any
"fifth" force field, and photon-paraphoton oscillations [5] could
resolve the discrepancy between the observed cosmic background and
pure blackbody radiation [6]. The discovery of gravitons, tensor
particles with spin 2, or antigravitons, vector particles with
spin 1, would be particularly exciting and would be fundamental in
the construction of a quantum theory of gravitation.

For the detection of massless or nearly massless weakly
interacting light vector bosons we propose a novel experimental
method based on coherent processes in refractive media. The
behaviour of visible photons in refractive media is determined by
Rayleigh scattering and a Feynman diagram of this process is shown
in Fig. 1a. Rayleigh scattering is an elastic process and the
total amplitude for the process is a coherent sum of scattering
amplitudes on many scattering centers. In addition to the Rayleigh
scattering there is also an elastic process in which a photon is
absorbed and a weakly interacting light vector boson is emitted. A
Feynman diagram for this process is shown in Fig. 1b. Weakly
interacting light spin 1 particles are denoted with $\xi$. In
refractive media the scattering amplitudes for this process also
add coherently and the weakly interacting vector bosons behave
like photons in the sense that they are emitted in both directions
at the interface between two refractive media, the backward
emission corresponding to reflection and the forward emission to
refraction.

We would like to calculate the intensities of reflected and
refracted beams for the processes shown in Fig. 1. If ~$a_{fi}$~
is the scattering amplitude for a single scatterer placed at the
origin of a coordinate system then the scattering amplitude for a
single scatterer at an arbitrary position $\vec{r}$  will be
$a_{fi}(\vec{r})=a_{fi}\exp(-i\vec{q}\vec{r})$, where
$\vec{q}=\vec{k}_i-\vec{k}_f$  is a vectorial change of the wave
vector. If the scatterers are densely packed so that the average
distance  $d$  between two scatterers satisfy the condition $qd\ll
1$, the total amplitude could be obtained by integrating
contributions from all scatterers,
\begin{equation}\label{eq:1}
   A_{fi}=a_{fi}\int d^3 r\mathcal{N}(\vec{r})\exp(-i\vec{q}\vec{r})
\end{equation}
In relation \reffz{eq:1} $\mathcal{N}(\vec{r})$ is a realistic
distribution of the scatterer number density. For two refractive
media with well defined scatterer number densities $N$ and $N'$,
$\mathcal{N}(\vec{r})$ could be represented with the well-known
step function $\theta$. If the plane surface between two media is
in the $x,y$ plane, than the density distribution
$\mathcal{N}(\vec{r})=N\theta(-z)+N'\theta(z)$ acquires constant
values $N$ for $z<0$, and $N'$ for $z>0$. By integrating relation
\reffz{eq:1} and using well known representation of $\theta$
function
%\end{document}
\begin{equation}\label{eq:2}
   \theta(\alpha)=\frac{1}{2\pi i}\int_{-\infty}^{+\infty}
   \frac{\exp(-i\alpha\tau)}{\tau-i\varepsilon} d \tau
\end{equation}
we obtain the result
\begin{equation}\label{eq:3}
    A_{fi}=a_{fi}i(2\pi)^{2}\delta(q_{x})\delta(q_{y})\frac{N-N'}{q_{z}}
\end{equation}
Wave numbers $k$ and $k'$  must be evaluated for particles in $N$
and $N'$ refractive media, and not for particles in free space.
For particles of energy $\omega$ in media $N$ the wave number is
$k=(\omega/c)n$, where $n=1+N a_{fi}(0^{o})/k_{i}^{2}$ is index of
refraction, $a_{fi}(0^{o})$ is the forward scattering amplitude
and $k_i$ is the wave number of particles in free space. Eq.
\reffz{eq:3} does not depend on the type of particles involved in
the scattering processes and will be the basis of all our further
investigations.

For photons we can obtain the reflection and Snell refraction laws
and also the reflection and refraction coefficients
$P^{refl}_{\gamma\rightarrow\gamma}$ and
$P^{refr}_{\gamma\to\gamma'}$. We can extend this to the situation
where photons incident from medium N scatter at the interface
between the $N$ and $N'$ media, and both photons and weakly
interacting vector bosons are emitted. For the process in which
weakly interacting vector bosons are emitted into medium $N'$, in
analogy with the similar process in conventional electrodynamics,
we define the vector boson refraction coefficient
\begin{equation}\label{eq:4}
   P^{refr}_{\gamma\to\xi'}=\frac{k_{\xi' z}\mid
   A^{refr}_{\xi'\gamma}\mid^2}{k_{\gamma'z}\mid
   A^{refr}_{\gamma'\gamma}\mid^2+k_{\gamma z}\mid{A^{refl}_{\gamma\gamma}\mid^2}+k_{\xi'
   z}\mid{A^{refr}_{\xi'\gamma}\mid^2}+k_{\xi
   z}\mid{A^{refl}_{\xi\gamma}\mid^2}}
\end{equation}
In Eq. \reffz{eq:4} $A^{refl}_{\gamma\gamma}$ and
$A^{refr}_{\gamma'\gamma}$  are the total amplitudes for the
reflection and refraction of the photon beam, and are obtained
from Eq. \reffz{eq:3} by inserting ${q_z}={2k_\gamma}
\cos{\theta_i}$ and
$q_z=k_{\gamma}\cos{\theta_i}-k_{\gamma'}\cos{\theta}_{\gamma'}$,
respectively. $\theta_i$ and $\theta_{\gamma'}$ are the angles of
incident and refracted photon beams defined relatively to the
normal on the plane surface between two media. $k_\gamma$ and
$k_{\gamma'}$ are wave numbers of photons in $N$ and $N'$
refractive media, and  $k_{\gamma z}$ and $k_{\gamma'z}$ are their
projections on the $z$-axis. The total scattering amplitudes
$A^{refl}_{\xi\gamma}$ and $A^{refr}_{\xi'\gamma}$ for the
processes in which weakly interacting vector bosons are emitted
into $N$ and $N'$ refractive media could be obtained from Eq.
\reffz{eq:3} by inserting $q_z={k_\gamma}{\cos\theta_i}-{k_\xi}
{\cos\theta_{\xi}}$ and
$q_z={k_\gamma}{\cos\theta_i}-{k_{\xi'}}{\cos\theta_{\xi'}}$,
respectively. $k_\xi$ and $k_{\xi'}$ are wave numbers of weakly
interacting vector bosons emitted at $\theta_\xi$ and
$\theta_{\xi'}$ angles into $N$ and $N'$ refractive media, and
$k_{\xi z}$ and $k_{\xi' z}$ are their projections on the
$z$-axis.

The main results of our investigations will become clearer if we
make some approximations that will simplify Eq. (4). First we will
assume that the rest masses of the weakly interacting vector
bosons are very small compared to their total energies. If we
further assume that the angle of incidence $\theta_i \approx 0^o$
we eliminate the dependence of scattering amplitudes $a_{fi}$ on
initial polarization and scattering angles. For the ratio of
relevant scattering amplitudes we can make a reasonably good
approximation $|a_{\xi\gamma}|^2/|a_{\gamma\gamma}|^2 \approx
{\alpha_\xi}/{\alpha<<1}$. In that case from Eqs. (3) and (4) we
obtain
\begin{equation}\label{eq:5}
   P^{refr}_{\gamma\to{\xi'}}\approx\left(\frac{\alpha_{\xi}}{\alpha}\right)
   \frac{4n_\gamma {n_{\xi'}}} {(n_\gamma-n_{\xi'})^2}
   \left(\frac{n_{\gamma'}-n_\gamma}{n_{\gamma'}+n_\gamma}\right)^2
\end{equation}

where $n_\gamma$ and $n_{\gamma'}$ denote indices of refraction
for photons, and $n_{\xi'}$ for new vector bosons, respectively;
$\alpha_\xi$ and $\alpha$ are dimensionless constants of
interaction for new vector bosons and photons with electrons,
respectively.

{} From Eq. \reffz{eq:5} we see that if photons were incident from
the high vacuum region, the number of light vector bosons produced
in the scattering process at the surface between two refractive
media could be very high, because $n_{\xi'}\approx 1$ and
therefore $P^{refr}_{\gamma\to{\xi'}}\propto 1/({n_\gamma}-1)^2$,
and for high vacuum $n_{\gamma}\approx 1$. On the other hand the
intensities of new vector bosons emitted in the reflection
processes are always by an approximate factor $\alpha_\xi/\alpha$
lower than the intensities of the reflected photons. Therefore we
would be most interested in refraction processes in which photons
are absorbed and new vector bosons are emitted in the forward
direction. Sensitivity of our experimental method is limited by
the masses of light vector bosons produced in the scattering
process, because for non-negligible masses the change of the
$z$-component of the wave vector is $q_z={k_{\gamma z}}-{k_{\xi'
z}}\approx \left[{n_\gamma}-1+(m_{\xi}c^2/E_\gamma)^2/2 \right]$
for $\theta_i\approx0^o$ scattering.

For the experimental investigations of weakly interacting light
vector bosons we propose the "shining light through the wall" type
of experiment. The initial photon from the laser is incident from
the high vacuum region and scatters at a plane surface of a thin
glass plate with an index of refraction $n_{\gamma'}$  for
photons. The photon beam is then blocked to eliminate everything
except the light vector bosons, which pass through because of
their extremely weak interaction with ordinary matter. The light
boson then interacts at the surface between another glass plate
and the high vacuum region to produce a photon, whose detection is
the signal for the production of the light boson. A single photon
counting system using a low noise photomultiplier could be very
efficient for the detection of photons. The sensitivity of the
system can be increased if we use $L$ and $L'$ glass plates in
front and behind of the blocking piece. In order to minimize the
absorption of photons in the glass, the plates should be tilted at
the Brewster angle. The expected number of photons recorded in the
single photon counting system will be
\begin{equation}\label{eq:6}
   N_c=\varepsilon L P^{refr}_{\gamma\rightarrow{\xi'}}L'
   P^{refr}_{{\xi'}\rightarrow\gamma}N_{laser}
\end{equation}

where $\varepsilon$ is the efficiency of the photomultiplier,
$N_{laser}$ is total number of photons emitted from the laser
during the experiment, and $P^{refr}_{{\xi'}\rightarrow\gamma}$ is
the coefficient for the refraction process, given by an expression
similar to Eq. (5), in which weakly interacting vector bosons
produce photons.

To estimate the sensitivity of our experiment we will assume that
the pulsed laser emits the $532$\, nm wavelength photons with the
total energy of $30$\, mJ per $10$\, nsec width pulse at the
repetition rate of $20$\, Hz. We will further assume
$\varepsilon=0.25$, $L=L'=1000$,  $n_{\xi}= n_{\xi'}=1$,and
$n_{\gamma'}=1.9$. Index of refraction $n_\gamma$ will be
expressed as the function of residual pressure in the experimental
apparatus. Using coincidence techniques we can expect the
background to be less than $10$ counts during $10^6$ seconds of
laser operation and then from Eqs. (5) and (6) we easily calculate
limits for $\alpha_\xi/\alpha$ at different vacuum values. Results
of our estimates are shown with the full line in Fig. 2.

Present limits for several weakly interacting vector bosons are
also shown in Fig. 2. For even modest vacuums of $\approx 10^{-8}$
Torr the upper limits on coupling constants for paraphotons [7]
with rest masses of $m_{\Gamma}\leq 10^{-7}$ eV could be lowered
by an enormous factor of $\approx10^{28}$. For very high vacuums
of $\approx 10^{-12}$ Torr, which are still available in
earth-bound laboratories, the upper limit on coupling constants
for leptonic photons [8] could be improved by a factor of $\approx
10^4$.

We have also made an estimate for detecting some possible
gravitational effects. If a gravity theory would allow the
existence of a vector field then antigravitons [9], i.e. spin 1
gravitational field quanta, could be detected in our apparatus.
The probability that the transition between the two electron
states will proceed by emission of gravitational, rather than
electromagnetic, radiation is typically of order
$\alpha_{grav}/\alpha\approx {Gm^2/e^2}$, where $G$ is
gravitational constant, and $m$ and $e$ are electron mass and
charge respectively. In that case antigravitons could be detected
even at vacuums of $\approx 10^{-8}$ Torr.

The authors gratefully acknowledge A. Per\v{s}in for many helpful
discussions concerning experimental realities, along with M.
Stip\v{c}evi\'{c} and B.A. Logan. Early assistence of B. Laki\'{c} is also
acknowledged. This work was supported by the Ministry of Science
and Technology of Croatia.

\newpage \vspace*{-7.5cm}  {\bf Figure captions}

\begin{figure}
\caption{Feynmam diagrams for the elastic scattering of photons on
atomic electrons. $\xi$ represents weakly interacting light
bosons. \label{fig1}}
\end{figure}

\begin{figure}
\caption{Sensitivity of the proposed method for different vacuum
values; corresponding masses of light bosons are shown on the
upper scale of the $x$-axis. Shaded regions are excluded by the
experiments. Exclusion region for paraphotons was obtained from
ref. [7]. Upper limit  on the leptonic photon coupling constant
was taken from ref. [8]. \label{fig2}}
\end{figure}

\newpage

\begin{center}
\mbox{\epsfxsize = 12 cm \epsfbox{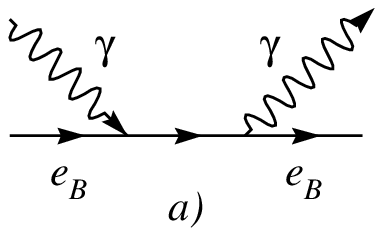}}
\mbox{\epsfxsize = 12 cm \epsfbox{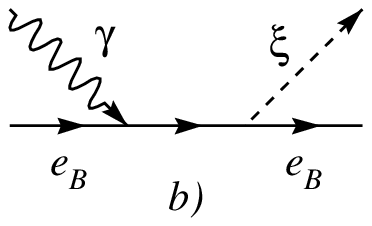}}
\end{center}

\begin{center}
\epsfxsize = 12cm \epsfbox{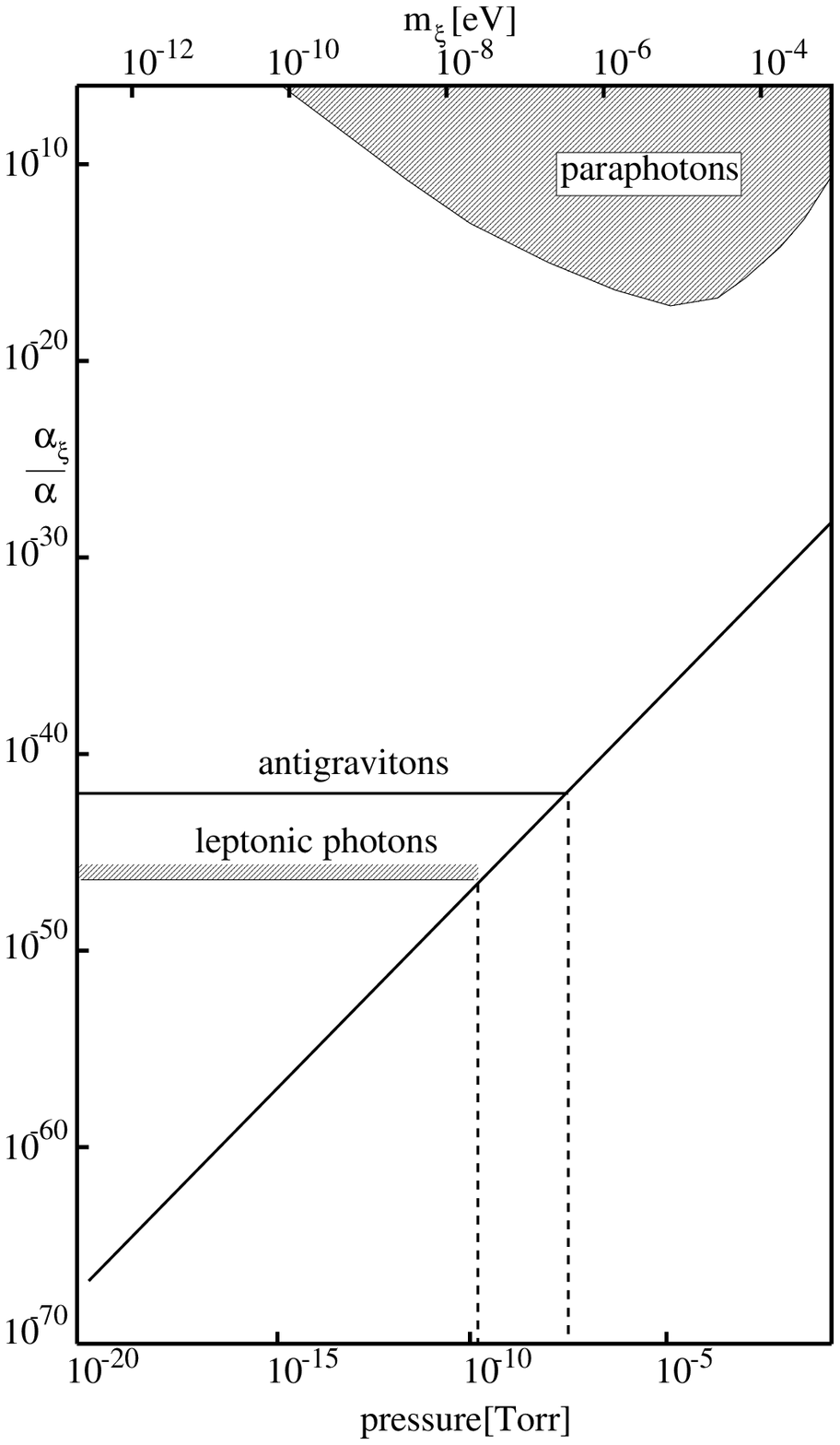}
\end{center}

\end{document}